\documentclass[conference]{IEEEtran}
\usepackage{hyperref}       
\usepackage{wrapfig}
\usepackage{subfig}
\usepackage{caption}
\usepackage{booktabs}       

\usepackage{cite}
\ifCLASSINFOpdf
  \usepackage[pdftex]{graphicx}
\else
\fi
\usepackage{amsmath,amssymb,amsthm}
\usepackage{amsmath,amsfonts,bm}

\def\eqref#1{equation~\ref{#1}}

\def\1{\bm{1}}

\DeclareMathAlphabet{\mathsfit}{\encodingdefault}{\sfdefault}{m}{sl}
\SetMathAlphabet{\mathsfit}{bold}{\encodingdefault}{\sfdefault}{bx}{n}

\usepackage{algorithm}
\usepackage{algorithmic}

\usepackage{array}
\usepackage{url}

\begin{document}
%
\title{Retrieval Augmented Generation Systems: Automatic Dataset Creation, Evaluation and Boolean Agent Setup}


\author{\IEEEauthorblockN{1\textsuperscript{st} Tristan Kenneweg}
\IEEEauthorblockA{\textit{Technische Fakultät} \\
\textit{University of Bielefeld}\\
Bielefeld, Germany \\
tkenneweg@techfak.uni-bielefeld.de}
\and
\IEEEauthorblockN{2\textsuperscript{st}Philp Kenneweg}
\IEEEauthorblockA{\textit{Technische Fakultät} \\
\textit{University of Bielefeld}\\
Bielefeld, Germany \\
pkenneweg@techfak.uni-bielefeld.de}
\and
\IEEEauthorblockN{3\textsuperscript{st}Barbara Hammer}
\IEEEauthorblockA{\textit{Technische Fakultät} \\
\textit{University of Bielefeld}\\
Bielefeld, Germany \\
bhammer@techfak.uni-bielefeld.de}
}


%


\maketitle

\begin{abstract}
  Retrieval Augmented Generation (RAG) systems have seen huge popularity in augmenting Large-Language Model (LLM) outputs with domain specific and time sensitive data. Very recently a shift is happening from simple RAG setups that query a vector database for additional information with every user input to more sophisticated forms of RAG. However, different concrete approaches compete on mostly anecdotal evidence at the moment. In this paper we present a rigorous dataset creation and evaluation workflow to quantitatively compare different RAG strategies. We use a dataset created this way for the development and evaluation of a boolean agent RAG setup: A system in which a LLM can decide whether to query a vector database or not, thus saving tokens on questions that can be answered with internal knowledge. We publish our code and generated dataset online. 
\end{abstract}


%
\IEEEpeerreviewmaketitle

\section{Introduction}
\label{sec:intro}

Large Language Models have seen huge progress in recent years and are able to generate coherent text on a wide variety of topics. The most recent example is GPT-4~\cite{brown2020language}, which is able to perform limited reasoning about the world~\cite{mitchell2023comparing}. However, LLMs lack domain specific and time sensitive data~\cite{kandpal2023large}. This is a problem for many real world applications. For example, a chatbot that is used by a project manager needs to know about company internal data and a chatbot that is used by a sports fan needs to know about the latest sports results.

To solve this problem, Retrieval Augmented Generation (RAG) systems have been proposed. In a RAG system a LLM is augmented with a vector database that contains relevant data. Depending on the user query relevant data is retrieved and injected into the LLM context. This approach has been shown to be very effective~\cite{lewis2020retrieval}. Many different RAG setups have been proposed, however they are mostly evaluated on anecdotal evidence at the moment.
In order to evaluate different RAG setups quantitatively a dataset is needed which is not contained in the initial LLM training set. Since modern LLMs are trained on large parts of the internet such data can be challenging to find. When evaluating 300 questions about random Wikipedia articles regarding truthfulness and relevance, for example, we find that GPT-4-0613 gets near perfect scores on most questions. See Figure~\ref{fig:baselines_a} for details. This happens because Wikipedia is contained within the training set. 

To address this problem we propose an automatic dataset creation workflow that can be used to generate datasets from Wikipedia articles and other sources and is suited for arbitrary LLM cutoff points and automatic evaluation. The datasets consist of matching questions and articles, wherein the article contains all information that is necessary to answer the questions. The questions are LLM generated. The articles are curated such that they mostly contain information about events post LLM cutoff point. 

Furthermore, we show how to implement an automatic evaluation to evaluate RAG systems on our dataset with regards to truthfulness and relevance. We base this automatic evaluation on the work of~\cite{liu2023gpteval} and~\cite{lin-chen-2023-llm}.

To test our dataset and evaluation workflow we investigate the boolean agent RAG setup. The naive RAG approach of querying the vector database with every user input query is very inefficient. If RAG is triggered at every user input, multiple pages of information are injected into the LLM context, even if the user input is a simple \textit{hello}. This is a waste of token usage and can be avoided by using a boolean agent RAG setup. In such a setup the LLM decides whether to call the information retrieval system for more data for each user input. This approach promises to generate comparable results while saving token usage on simple queries. In this paper we present a working boolean agent RAG setup and give recommendations towards scenarios in which boolean agent RAG performs similarly to a naive RAG setup while saving tokens.

To summarize our key contributions are the following:

\begin{enumerate}
    \item We present a dataset creation workflow designed for evaluating RAG systems. This workflow enables the generation of datasets from Wikipedia articles and other sources. The datasets can be configured to predominantly contain information beyond an LLM's cutoff point, and are suitable for automatic evaluation. We make the datasets we created publicly available.
    \item We show how to perform automatic evaluation on our datasets.
    \item We use a dataset created by our workflow for the development and implementation of a boolean agent RAG setup.
    \item We give recommendations under which circumstances a boolean RAG setup can be deployed to save tokens while maintaining performance.
\end{enumerate}

We make our code and datasets publicly available at \href{https://github.com/TKenneweg/RAG_Dataset_Gen}{www.github.com/TKenneweg/RAG\_Dataset\_Gen}.

\section{Related Work}
\label{sec:Background}

Retrieval augmented generation was first proposed by Lewis et al.~\cite{lewis2020retrieval}. They enhance a seq2seq model with a non-parametric dense vector index of Wikipedia and outperform multiple task-specific state-of-the-art systems. 
RAG in the context of LLMs works by embedding text chunks into a vector space. Inside the vector space similar text chunks are close to each other and can be found using fast nearest neighbor search. The text chunks which are near in vector space to a user query are injected into the LLM context. The LLM can use the chunk information to generate a better and more factually grounded answer.
Lewis et al.\ used BERT~\cite{DBLP:journals/corr/abs-1810-04805} as an embedding model. In 2022 OpenAI released Ada-002~\cite{openai2022embedding} which is a state-of-the-art embedding model and has been shown to outperform other embedding models on multiple tasks such as text search, code search, sentence similarity and text classification.
A further central element of RAG is chunking which refers to the process of splitting a text into multiple chunks, which subsequently get embedded into the vector database. Different chunking strategies have been proposed ranging from simple sentence chunking to more complex algorithms that rely on domain specific information. An often used generic form of chunking is recursive chunking, which recursively splits a long text into smaller chunks using a list of separators like newline characters until chunks of the required size are reached~\cite{finardi2024chronicles}. 

There are many methods to augmented RAG systems. Two of the most popular are Hyde retrieval~\cite{gao2022precise} and RAG with guardrails~\cite{RagGuardrails}. Hyde Retrieval augments the retrieval process by querying a LLM to write a hypothetical document which contains the relevant data. This hypothetical document is then embedded and used to query the vector database. Since the hypothetical document is presumably nearer in embedding vector space to the relevant data than the user query this can lead to better retrieval results. 
Rag with guardrails checks if the embedding vector of a user query is within a certain region of the vector space. This region has been specified beforehand by embedding a set of example queries. These example queries can be positive or negative. Depending on the desired system behavior data retrieval can be rejected if the user query embedding falls within a wrong part of the embedding space.
As many different methods of RAG system augmentation exist, it is important to have the tools to evaluate them quantitatively. 

Automatic evaluation of LLM output has become an active subject of recent research, since manual labeling is infeasible in many cases and limits the iteration speed of research drastically. Liu et al.~\cite{liu2023gpteval} propose G-EVAL, a method to automatically evaluate LLM output with regards to truthfulness, relevance and fluency. They use GPT-4 to evaluate different metrics of a LLM given answer and use different prompting techniques to achieve this. They compare the LLM evaluations to human generated labels and conclude that strong correlations exist. Lin et al.~\cite{lin-chen-2023-llm} propose LLM-EVAL, which utilizes a new prompting technique to generate scores for multiple dimensions like truthfulness and relevance in one prompt.

The automatic generation of language datasets is a very new discipline that has emerged with the advent of LLMs. In~\cite{dai2022promptagator} Zhuyun et al.\ show how to generate large amounts of synthetic data from few example data to create task specific retrievers. RAGAS~\cite{es2023ragas} is a system that focuses on evaluating RAG systems regarding the match between generated answer and provided context, by defining three different metrics that are evaluated at the sentence and statement level. They also employ gpt-3.5 to generate questions about 50 Wikipedia articles, but do not supply a baseline evaluation of the generated questions. In contrast, we focus on the dataset creation process and provide an automatic evaluation method along the lines of G-EVAL~\cite{liu2023gpteval} and LLM-EVAL~\cite{lin-chen-2023-llm}.

\section{Dataset \& Evaluation Workflow}
\label{sec:methods}
The evaluation of integrated LLM systems is in its early stages and no best practices exist for the evaluation of RAG systems. Therefore, we put great effort into designing a dataset creation and automatic evaluation workflow that stands up to rigorous scrutiny.

\subsection{Dataset}
\begin{figure*}
  \vspace{-0.05\textwidth}
  \subfloat[$A_r$]{\includegraphics[width = 0.49\textwidth]{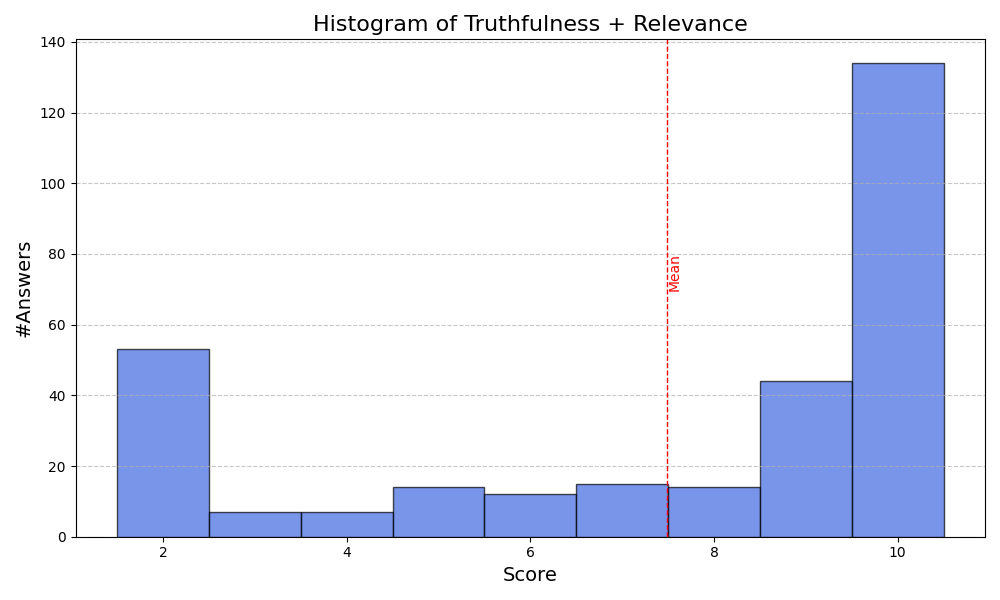}
  \label{fig:baselines_a}}
  \subfloat[$A_d$]{\includegraphics[width = 0.49\textwidth]{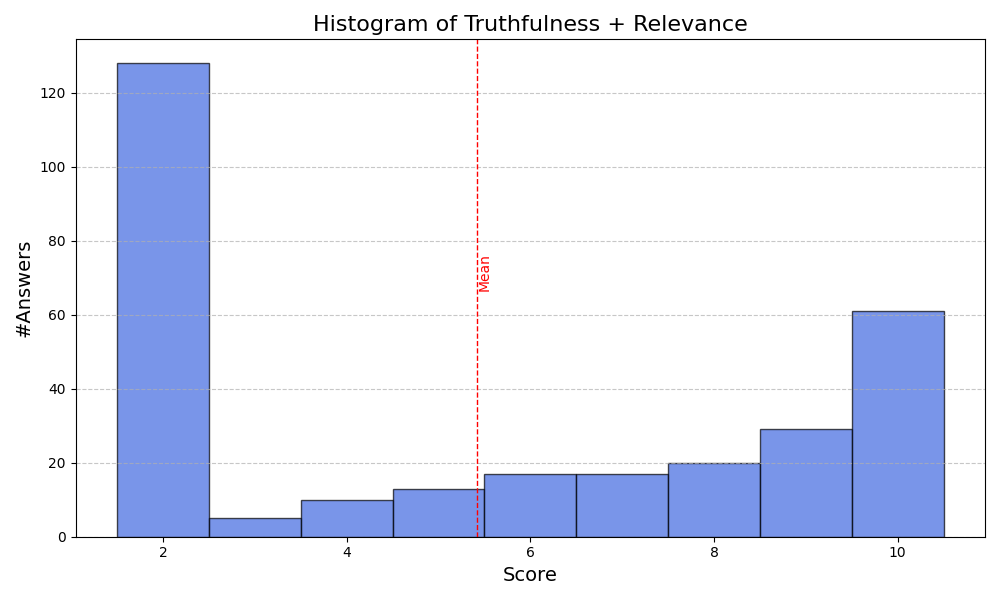}
  \label{fig:baselines_b}}\\
  \subfloat[$A_f$]{\includegraphics[width = 0.49\textwidth]{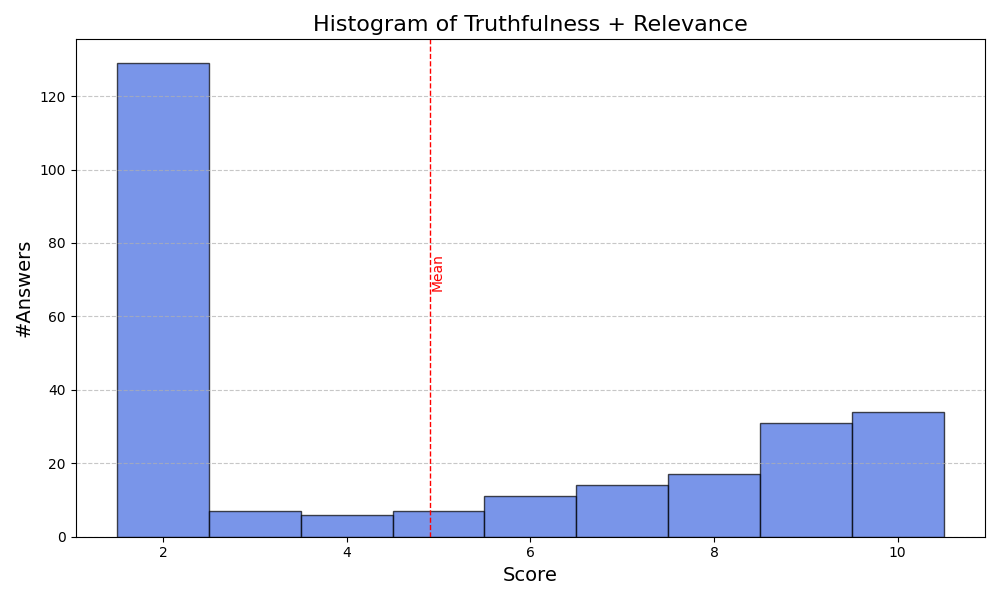}
  \label{fig:baselines_c}} 
  \subfloat[$A_{f}$ GT]{\includegraphics[width = 0.49\textwidth]{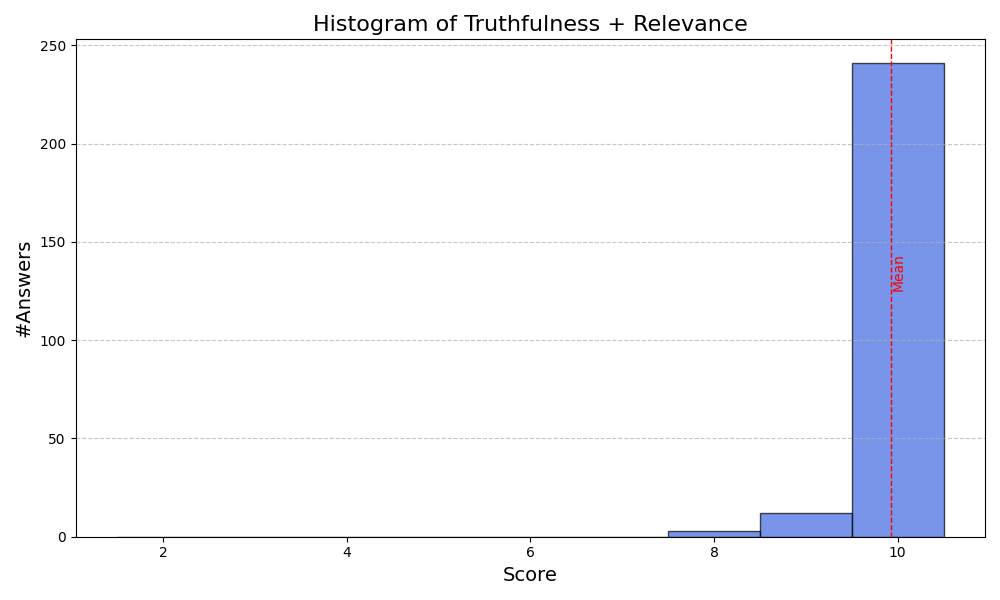}
  \label{fig:baselines_d}}

  \caption{Sum of accuracy and relevance for different baseline test setups. a) no RAG 300 random articles from $A_r$, b) no RAG 300 articles from $A_d$, c) no RAG all 256 articles from $A_f$, d) $A_f$ with the correct article supplied to the answerer.}
  \label{fig:baselines}
  \vspace{-0.01\textwidth}
  \end{figure*}

To evaluate the effectiveness of a RAG setup we need a challenge in which a LLM profits from the supplied RAG information. The suitability of generic Wikipedia question answering is limited. We show this by downloading 300 random Wikipedia articles. We generate creative questions for each article using GPT-4 with a high temperature for maximum variability. Afterwards we let GPT-4 answer the questions without the corresponding articles in context. Figure~\ref{fig:baselines_a} shows the results. As we can see a substantial amount of questions (127) are answered perfectly without any additional information. This is due to the fact that Wikipedia is contained within the training set.

To solve this problem we propose the following workflow to create an article-question dataset from Wikipedia, specifically for RAG evaluation:

\begin{enumerate}
    \item Download $n_r$ random Wikipedia articles. Filter for those that have been created after the knowledge cutoff date of the LLM that is tested. For GPT-4-0613 this cutoff date is September 2021, so we collect articles from October 2021 onward. We denote the first 300 of these $n_d$ articles as $A_d$.
    \item Use GPT-4 to give a yes no answer if the majority of information in an article is about a topic that happened after the cutoff date. This may include politics or sports in that time period but also for example new computing libraries that have been published since. Continue with $n_f$ articles about new topics. In our case $n_f=256$. We denote the set of these articles as $A_f$.
    \item Generate one or more questions per article using GPT-4 with a high temperature for maximum variability. 
\end{enumerate}

This process can be used to crate a dataset of arbitrary size. All questions are associated with an article that can be served to the evaluating network as a means of ground truth. We download $n_r=12792$ random articles, which we filter for date to get $n_d=818$ articles. We then use GPT-4 to classify the articles and get $n_f=256$ articles.

We manually assess the quality of the generated questions and find that the majority of questions are of high quality. Some examples are:

\begin{itemize}
	\item \textit{What are the three venues across Oxfordshire, England from which Hinksey Sculling School operates, and which age groups train at each venue?}
	\item \textit{What were the main criticisms of Manuchehr Kholiqnazarov's trial after his arrest in Tajikistan in 2022?}
	\item \textit{What is the title of the short story from Philip Fracassi's 2021 collection, Beneath a Pale Sky, that was optioned for a feature film adaptation in 2022?}
\end{itemize}

To test our dataset we ran baseline tests without RAG. Firstly, we ran a baseline test with 300 random articles, denoted as $A_r$. Next we ran a baseline test with the 300 articles from $A_d$ and all 256 articles from $A_f$. And finally we ran a ground truth baseline test on $A_f$ for which we supplied the correct article to the answerer. The results can be seen in Figure~\ref{fig:baselines}. As we can see the results are worst for $A_f$ which shows that our dataset is successfully filtered to include less question about topics that are present in the LLM training set. Furthermore, we get near perfect results of 4.96 for truthfulness and relevance on $A_f$ if we supply the correct article to the answerer. This indicates that our evaluation workflow is capable of determining truthfulness and relevance and that the article content does indeed contain the information to answer the given questions. Table~\ref{table:1} shows the average truthfulness and relevance for all baseline runs.

\subsection*{Dataset Analysis}

It is notable that the answer quality for $A_d$ is better than expected. We investigated this further. Manual perusal of Wikipedia articles within $A_d$ revealed that those articles contain a lot of information about events before the cutoff date. This is the case because while we can filter for creation date the contents of these articles are often about topics that are not recent. Some articles are for example about films which have been shot decades ago but only recently gotten an English Wikipedia entry. 

Furthermore, we note that the drop in answer quality from $A_d$ to $A_f$ is also smaller than might be expected. The GPT powered filtering step reduces the number of articles in $A_d$ from $n_d=818$ to $n_f=256$. Manual perusal confirms that the majority of information of the articles in $A_f$ is about recent events. However, we find two question types that allow GPT-4-0613 to generate good answers despite its lack of knowledge. One is common sense questions. An example:
\newline
\newline
\textit{Question: What does Hayley Kiyoko mean when she says the "color palette" of her second studio album, Panorama, is "darker" than her debut album, Expectations?}
\newline
\newline
\noindent Even without knowing who Hayley Kiyoko is or what her albums are about, a good guess at an answer can be made. The other question type is about articles that contain mostly recent information but deal with topics that are not recent. An example: 
\newline
\newline
\textit{Question: What role has social media played in the ZouXianZouxian phenomenon and the increase of Chinese migration through the US southern border?}
\newline
\newline
\noindent The article related to this question contains mostly information about the ZouXianZouxian phenomenon in the times of Covid. However, the phenomenon itself is not recent and GPT-4-0613 can answer the question to satisfaction using its internal knowledge. 

If a dataset is necessary which generates worse performance on baseline tests than $A_f$ this could presumably be achieved by modifying the question generation prompt to focus on recent topics. Our question generation prompt is given by:
\newline
\newline
\textit{Generate a creative question about the contents of the following text: \{text\}.}

\begin{table} 
	\centering
	\begin{tabular}{c cccccc}
		Metric & $A_r$ & $A_d$ & $A_f$& $A_f GT$ \\ 
		\cmidrule(r){1-1}   \cmidrule(r){2-5} 
		\it{Truthfulness} & 3.59 & 2.62 & 2.48 & 4.96\\ 
		\it{Relevance} & 3.9  & 2.80 & 2.42 & 4.96 \\  
		\cmidrule(r){1-1}   \cmidrule(r){2-5} 
		
	\end{tabular}
	\caption{Average truthfulness and relevance of the GPT-4-0613 generated answers to questions about Wikipedia articles. $A_r$ denotes 300 random articles, $A_d$ denotes 300 articles that have been created after the knowledge cutoff date of GPT-4-0613, $A_f$ denotes 256 articles that have been classified by GPT-4-0613 as mostly containing information about things after the knowledge cutoff date. $A_f GT$ denotes the same articles as $A_f$ but with the correct article supplied to the answerer.}
	\label{table:1}
 
\end{table}

\subsection{Automatic Evaluation}

In the past year two methods have seen larger adoption in the context of automated LLM evaluation: G-EVAL~\cite{liu2023gpteval} and LLM-EVAL~\cite{lin-chen-2023-llm}. Both use similar approaches. They use GPT4 to evaluate different metrics like truthfulness and relevance of a LLM given answer and use different prompting techniques to achieve this. 

We follow this previous work in spirit, however we decided to switch to the GPT4 function calling API to generate scores, since this increases the output reliability. We rate truthfulness and relevance on a score of 1 to 5 and omit fluency, since it is no longer a concern for modern LLM systems.
We set the general behavior in the system prompt:
\newline
\newline
\textit{Your task is to evaluate answers given by a chatbot. You are provided the user query, the chatbot generated answer and a wikipedia article that contains information about the true answer.
Given this information generate two scores from 1 to 5, where 5 is the best, for the chatbot generated answer,
concerning relevance and truthfulness. Give a score of 1 for relevance if the answer is that the chatbot doesn't know.}
\newline
\newline
\noindent and design a small and simple function calling description:
\newline
\newline
\noindent \textit{Set the answer evaluation for truthfulness and relevance.}
\newline
\newline
\noindent With similar description for the parameters \textit{truthfulness} and \textit{relevance}.

A further problem to address when automatically evaluating LLM answers with regards to truthfulness is the symmetry in knowledge between question answerer and evaluator. By design our dataset contains question-article pairs in which the article contains relevant information to answer the question. The correct article is passed to the evaluator but not the answerer, unless this happens implicitly by the RAG system design. 

\section{Boolean Agent RAG Evaluation}

Having established a dataset creation and evaluation workflow we now turn to the evaluation of RAG setups using our created datasets.
The number of possible RAG configurations to tests is vast. We therefore focus on what we think is a configuration that is applicable to most real world scenarios: Boolean Agent RAG (BARAG). With BARAG we refer to a setup in which for each user input the LLM decides if it needs to query the vector database in order to answer. Compared to naive RAG this setup has the potential to save a large amount of tokens, since most tokens are not spent on user input and generated answers but rather the injected text which has been retrieved from the vector database and can be several pages long, even for simple queries. Thus, BARAG is a potential candidate for nearly all real world applications of RAG. 

\subsection*{Naive RAG}

\begin{figure}
	\begin{center}
	\includegraphics[width=\linewidth]{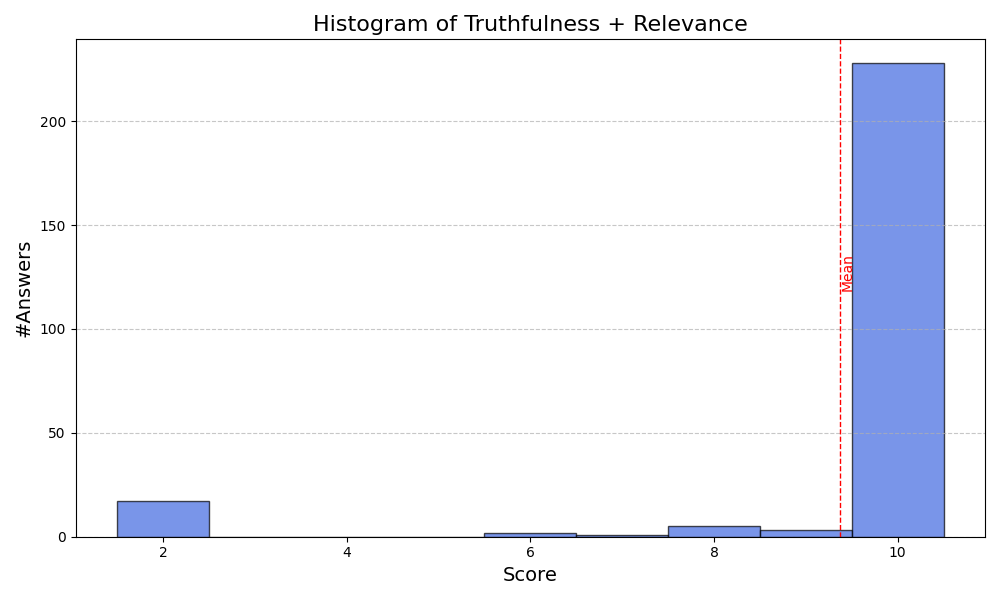}
	\caption{Results of using naive RAG on $A_f$. The average truthfulness is 4.71 and average relevance is 4.66.}
	\label{fig:naiverag}
\end{center}
\end{figure}

Before we commence to BARAG we set up a naive RAG system. There are many choices to make, such as chunk size and embedding method. The choices we make here are used in all further RAG tests. 

For the database we use our collection of $n_r=12792$ unfiltered random Wikipedia articles. We choose such a large database to simulate the challenging conditions of most real world RAG setups. For the embedding method we use OpenAIs Ada-002 model, as it seems to be the state-of-the-art embedding model at the time of writing. As distance metric for the vector database we use cosine similarity. We return 5 chunks for each vector database query.
For chunking we use recursive chunking with a maximum chunk size of 1024 characters and a maximum overlap of 48 characters. Additionally, we add the title of each Wikipedia page to every chunk that is generated from it to ensure that chunk information is not without context. 

The results of using naive RAG on $A_f$ can be seen in Figure~\ref{fig:naiverag}. The average truthfulness is 4.71 and average relevance is 4.66. This shows that naive RAG is very effective on our dataset, especially when compared to the baseline of answering without RAG as seen in Figure~\ref*{fig:baselines_c}. 

However, it is also wasteful with regards to token usage. To answer the 256 questions of $A_f$, $n_{in} = 224319$ input tokens and $n_{out}= 24356$ output tokens were used. Text is injected into the LLM context for every query, even if the user input were a simple \textit{hello}, which is never the case here but presumably in many real world scenarios. 

\subsection*{Boolean Agent RAG}
Boolean agent RAG as we term it extends the naive RAG by a boolean decision step. Given the user input query the LLM decides if it needs to query the vector database. If so, it queries the vector database with the embedded user input query, if not it relies on its internal knowledge to answer the query. If this decision step is token efficient and works reliable it has the potential to save a large amount of compute. 

We start our investigation with a simple BARAG implementation. We use the OpenAI function calling API and design a function that accepts a boolean \textit{retrieve} as input. In the function calling description we pass the information that by setting this boolean to true a vector database is queried to provide additional information. Additionally, we add that this should only happen if necessary. 

We find that this approach does \textit{not} work. GPT-4-0613 decides to query the database nearly every time even on the $A_r$ dataset which contains over 127 perfectly scoring questions in the base test. We try to address the issue by prompt engineering without success. Even when adding -
\textit{If it is at all possible to answer a question without querying the database you should do so in order to save tokens} - to the prompt, GPT-4-0613 still queries the database 298 out of 300 cases on $A_r$. 
We hypothesized that due to the text completion background of the underlying LLM it answers the question in such a way as it excepts it to be true for the average person. We try to mitigate this by adding 
\newline
\newline
\textit{You are a knowledgeable AI that has been trained on large parts of the internet.}
\newline
\newline
to the system prompt, without effect. 

We conclude that GPT-4-0613 has insufficient awareness of its own capabilities to make a decision on when to need additional information in one true or false token. 

\subsection*{Advanced Boolean Agent RAG}
\begin{figure}
	\begin{center}
	\includegraphics[width=\linewidth]{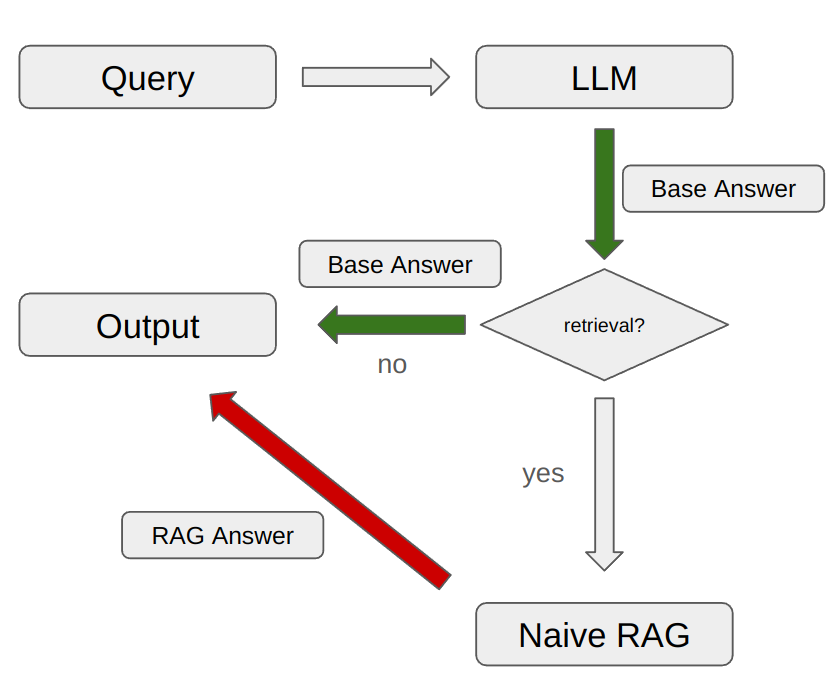}
	\caption{Schematic overview of the proposed boolean agent RAG system.}
	\label{fig:schema}
\end{center}
\end{figure}

\begin{figure*}
	\vspace{-0.05\textwidth}
	\subfloat[$A_r$]{\includegraphics[width = 0.49\textwidth]{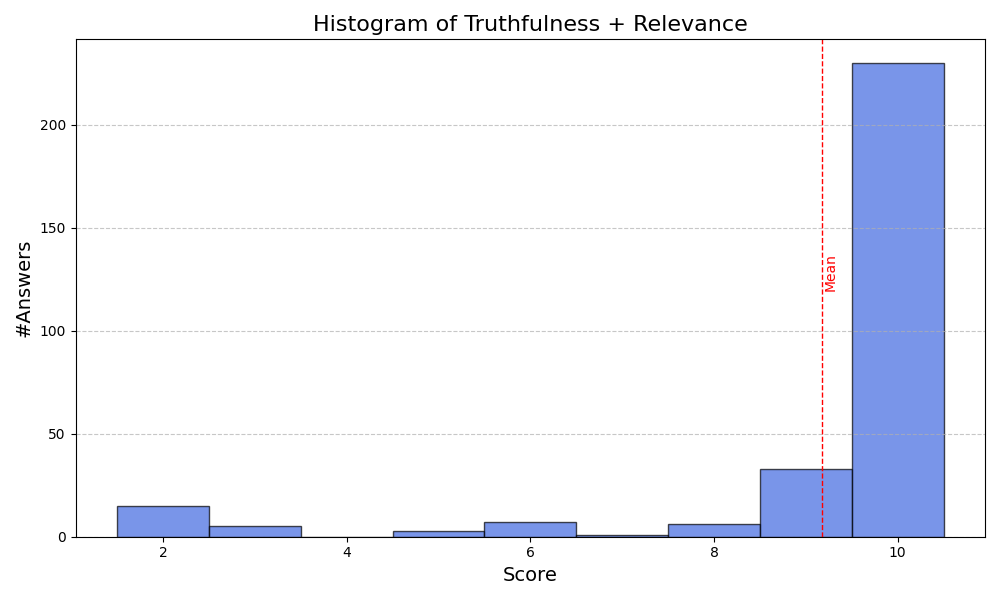}
	\label{fig:abarag_ar}}
	\subfloat[$A_f$]{\includegraphics[width = 0.49\textwidth]{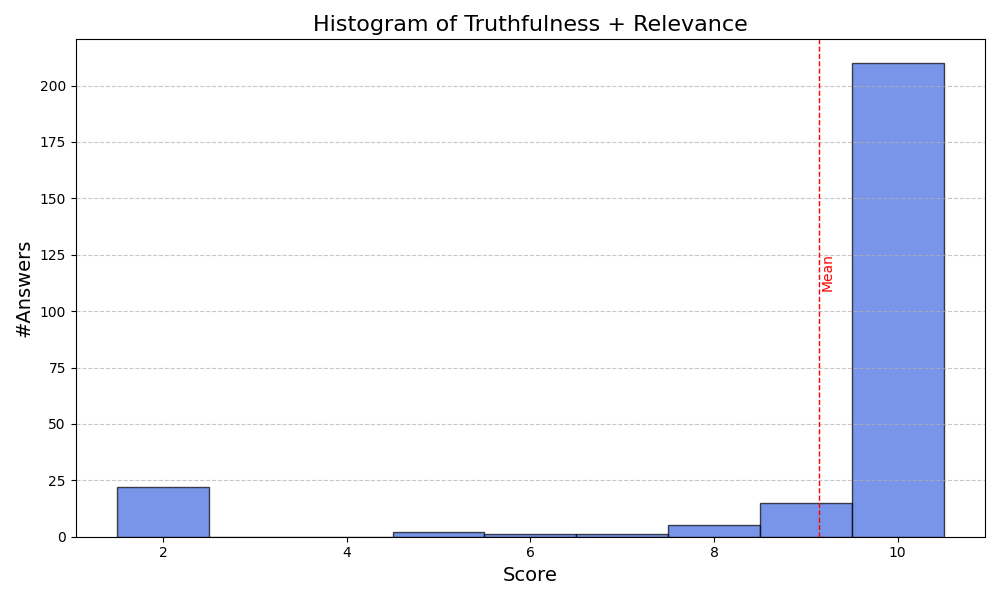}
	\label{fig:abarag_af}}
	\caption{Results of advanced boolean agent RAG on a) $A_r$, b) $A_f$.}
\label{fig:abarag}
\vspace{-0.01\textwidth}
\end{figure*}

\begin{table} 
	\centering
	\begin{tabular}{c cccccc}
		Metric & Baseline & NRAG & BARAG \\ 
		\cmidrule(r){1-1}   \cmidrule(r){2-4} 
		\it{Truthfulness} & 3.59 & 4.80 & 4.49\\ 
		\it{Relevance} & 3.9  & 4.80 & 4.67 \\  
		$n_{in}$ & 14k  &  262k & 224k\\  
		$n_{out}$ & 61k  & 34k & 78k \\  
		\it{\# retrieval} & 0  & 300 & 138\\  
		\cmidrule(r){1-1}   \cmidrule(r){2-4} 
		
	\end{tabular}
	\caption{Comparison of token usage and performance on $A_r$.}
	\label{table:a_r_token_usage}
	
\end{table}

\begin{table} 
	\centering
	\begin{tabular}{c cccccc}
		Metric & Baseline & NRAG & BARAG \\ 
		\cmidrule(r){1-1}   \cmidrule(r){2-4} 
		\it{Truthfulness} & 2.49 & 4.71 & 4.56\\ 
		\it{Relevance} & 2.42  & 4.66 & 4.59 \\  
		$n_{in}$ & 12k  &  224k & 260k\\  
		$n_{out}$ & 33k  & 24k& 57k \\  
		\it{\# retrieval} & 0  & 256 & 214\\  
		\cmidrule(r){1-1}   \cmidrule(r){2-4} 
		
	\end{tabular}
	\caption{Comparison of token usage and performance on $A_f$.}
	\label{table:a_f_token_usage}
	
\end{table}

To address the issues discussed above we extend the basic BARAG setup. In our new setup we let the LLM generate a baseline answer to the current user question and subsequently force a function call to decide if more information would have improved the answer. If the LLM decides that more information would have been beneficial we trigger database retrieval. We use a very simple function call description for this decision step: \textit{Set to true if you could have answered the last question better with more information}. A schematic overview of the system can be seen in Figure~\ref{fig:schema}.

We test this approach on $A_r$ and $A_f$ and compare the performance along with the token consumption in Tables~\ref{table:a_r_token_usage} and~\ref{table:a_f_token_usage}. Figure~\ref{fig:abarag} gives a more detailed performance overview.
We find that advanced boolean agent RAG triggers database retrieval in 138 out of 300 cases on $A_r$ and 214 out of 256 cases on $A_f$. We also see a slight decrease in answer quality for both datasets compared to naive RAG. For $A_f$ we see no drop in the number of input tokens. On the contrary we see an increase. This is reasonable, since $A_f$ is a dataset which is specifically designed to contain no questions that GPT-4-0613 can answer with its internal knowledge alone. Consequently, retrieval is triggered very often while we also have to pay for the token overhead generated by the base answer generation and subsequent function call. For $A_r$ we see a decrease in input tokens. This is due to the fact that $A_r$ contains many questions that can be answered with internal knowledge, which results in the low number of 138 retrievals. However, we also see a similar increase in output tokens, due to the overhead produced by base answer generation in the case where retrieval is triggered. 

All in all we cannot give a clear recommendation to use boolean agent RAG over naive RAG as a first solution in real world applications. However, we are confident that boolean agent RAG as described here can be used to save tokens under the right circumstances. Our datasets exclusively contains topic specific questions, while real world chat applications contain many filler phrases such as \textit{hello} or \textit{how are you} and many generic questions like \textit{What is the capital of Brazil?}, which an LLM can answer using its internal knowledge. In these cases boolean agent RAG might be used to save tokens.
Furthermore, better prompting techniques might increase the token efficiency of boolean agent RAG, while keeping comparable answer quality. In order to save cost the baseline answering and decision step could be handled by a less powerful and cheaper LLM like GPT-3.5. We publish our dataset and put this out as a challenge to the community.

\section{Conclusion}

In this paper, we present a comprehensive dataset creation workflow specifically tailored for the automated evaluation of (Retrieval-Augmented-Generation) RAG systems. Our workflow allows filtering for data beyond an arbitrary knowledge cutoff of a given LLM, thus allowing to create a dataset that is not contained within the internal knowledge of the LLM. Through comprehensive testing, we demonstrate the effectiveness of our dataset in evaluating RAG systems. Additionally, we explore the feasibility of implementing a boolean agent RAG system. Our findings reveal that a basic boolean agent RAG approach is ineffective. However, we have developed a sophisticated boolean agent RAG model capable of conserving tokens under certain conditions without compromising the quality of the generated responses. We propose the adoption of our dataset and evaluation methodology for future assessments of RAG systems. Furthermore, we challenge the research community to enhance boolean agent RAG configurations to optimize token conservation while preserving the quality of responses.


\section*{Acknowledgment}
The authors were supported by SAIL. 
SAIL is funded by the Ministry of Culture and Science of the State of North Rhine-Westphalia under the grant no NW21-059A.



\bibliography{references}
\bibliographystyle{IEEEtran}
%

\end{document}